\documentclass[final,5p,times,twocolumn]{elsarticle}

\usepackage[utf8]{inputenc}
\usepackage{textcomp} 
\usepackage{pifont}
\usepackage{graphicx}
\usepackage{xcolor}
\usepackage{booktabs}
\usepackage{threeparttable}
\usepackage{amssymb}
\usepackage{amsmath}
\usepackage{tabularx}
\usepackage{multirow}
\usepackage{caption}
\usepackage{adjustbox}
\usepackage{makecell}
\usepackage{stfloats}
\usepackage{placeins} 
\usepackage{url}
\usepackage{cuted}
\usepackage[linesnumbered,ruled,vlined]{algorithm2e}
\SetKwComment{tcp}{\%~}{}
\newcolumntype{C}{>{\centering\arraybackslash}X}

\biboptions{numbers,sort&compress}

\usepackage{hyperref}
\hypersetup{
    colorlinks=true,      
    linkcolor=blue,       
    citecolor=blue,      
    urlcolor=magenta     
}

\journal{Computers in Biology and Medicine}

\begin{document}

\begin{frontmatter}

%% --- TITLE ---
\title{Context-Aware Asymmetric Ensembling for Interpretable Retinopathy of Prematurity Screening via Active Query and Vascular Attention}

%% --- AUTHOR & ADDRESS BLOCK ---
\author[1]{Md. Mehedi Hassan}
\author[1,2]{\texorpdfstring{Taufiq Hasan\corref{cor1}}{Taufiq Hasan}}

\address[1]{m-Health Lab, Department of Biomedical Engineering, Bangladesh University of Engineering and Technology, Dhaka, Bangladesh}
\address[2]{Center for Bioengineering Innovation and Design, Department of Biomedical Engineering, Johns Hopkins University, Baltimore, MD, USA}

\cortext[cor1]{Corresponding author. \\
\begin{minipage}[t]{1.0\linewidth}
\vspace{1pt}
Emails: taufiq.hasan@jhu.edu, taufiq@bme.buet.ac.bd (T. Hasan) \\
Emails: m.hassan.bme@gmail.com (M. M. Hassan)
\end{minipage}}

%% --- ABSTRACT ---
\begin{abstract}
Retinopathy of Prematurity (ROP) is among the major causes of preventable childhood blindness. Automated screening remains challenging, primarily due to limited data availability and the complex condition involving both structural staging and microvascular abnormalities. Current deep learning models depend heavily on large private datasets and passive multimodal fusion, which commonly fail to generalize on small, imbalanced public cohorts. We thus propose the Context-Aware Asymmetric Ensemble Model (CAA Ensemble) that simulates clinical reasoning through two specialized streams. First, the Multi-Scale Active Query Network (MS-AQNet) serves as a structure specialist, utilizing clinical contexts as dynamic query vectors to spatially control visual feature extraction for localization of the fibrovascular ridge. Secondly, VascuMIL encodes Vascular Topology Maps (VMAP) within a gated Multiple Instance Learning (MIL) network to precisely identify vascular tortuosity. A synergistic meta-learner ensembles these orthogonal signals to resolve diagnostic discordance across multiple objectives. Tested on a highly imbalanced cohort of 188 infants (6,004 images), the framework attained State-of-the-Art performance on two distinct clinical tasks: achieving a Macro F1-Score of 0.93 for Broad ROP staging and an AUC of 0.996 for Plus Disease detection. Crucially, the system features `Glass Box' transparency through counterfactual attention heatmaps and vascular threat maps, proving that clinical metadata dictates the model's visual search. Additionally, this study demonstrates that architectural inductive bias can serve as an effective bridge for the medical AI data gap.
\end{abstract}

\begin{keyword}
Active query \sep attention \sep explainable AI \sep multimodal fusion \sep multiple instance learning \sep retinopathy of prematurity
\end{keyword}

\end{frontmatter}

%% --- GRAPHICAL ABSTRACT ---
\begin{strip}
  \noindent\begin{minipage}{\textwidth}
    \vspace*{-5em} % Aggressively pulls the block up
    \centering
    \bfseries\small G\hspace{0.5em}R\hspace{0.5em}A\hspace{0.5em}P\hspace{0.5em}H\hspace{0.5em}I\hspace{0.5em}C\hspace{0.5em}A\hspace{0.5em}L\hspace{2.5em}A\hspace{0.5em}B\hspace{0.5em}S\hspace{0.5em}T\hspace{0.5em}R\hspace{0.5em}A\hspace{0.5em}C\hspace{0.5em}T \\[0.1em]
    
    \vspace{-0.5em} % Tightens space between text and line
    \rule{0.7\linewidth}{0.5pt} \\[0.5em] % Line is now same width as image
    
    % Reduced width to 0.7 to fit on the first page
    \includegraphics[width=0.7\linewidth]{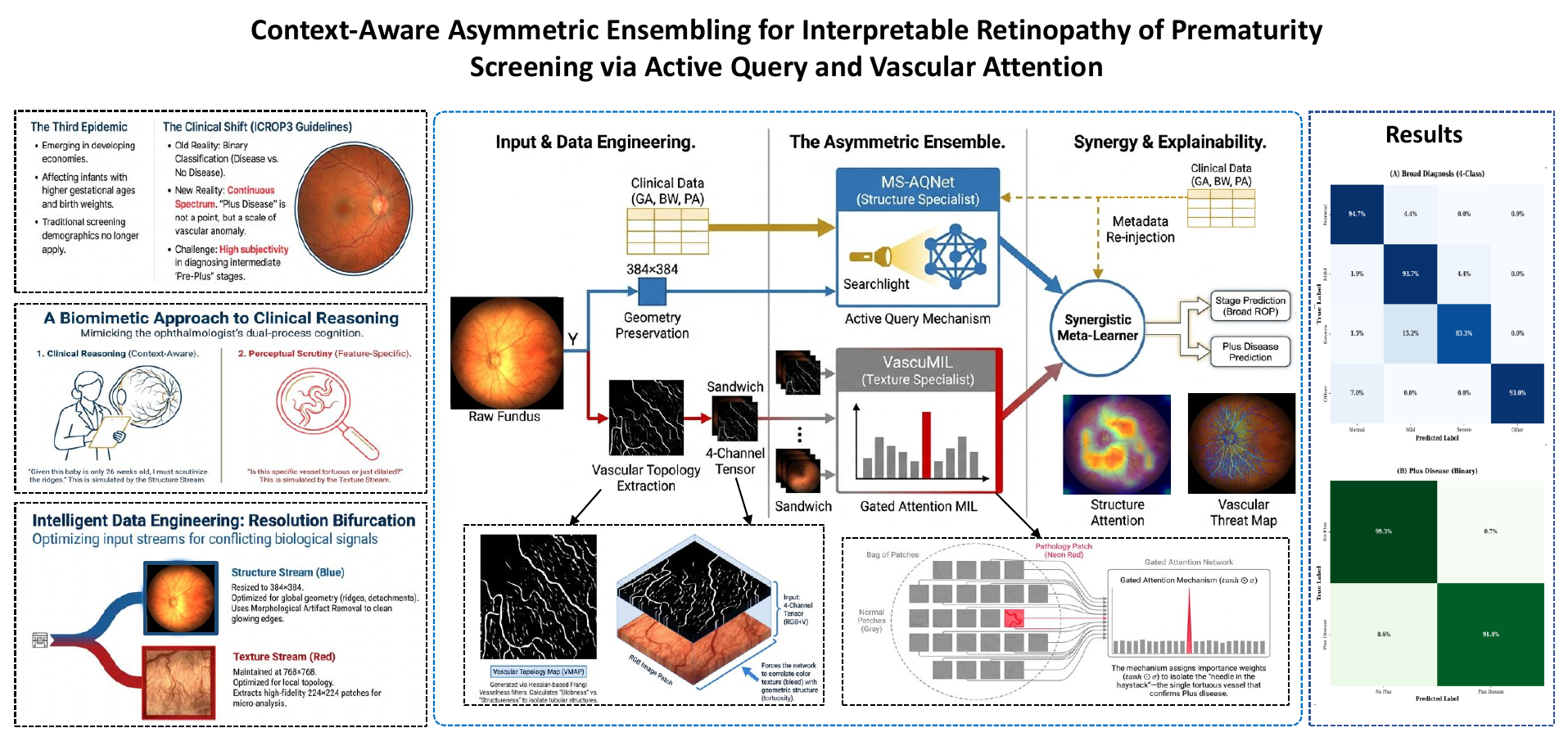}
    \vspace{-1em} % Pulls the main text up closer to the image
  \end{minipage}
\end{strip}

\section{Introduction}
\label{sec:introduction}
Retinopathy of Prematurity (ROP) continues to pose a daunting challenge to pediatric ophthalmologists and is a prime cause of preventable blindness in childhood. Although there has been a decrease in severe ROP incidence in developed countries, a `third epidemic' of ROP is expanding within middle-level developing economies\cite{hong_retinopathy_2022}. Variations in neonatal care have changed ROP epidemiology, and unlike before, treatment-requiring ROP in regions like India and Turkey is now often seen within higher gestational ages and birth weight infants\cite{vinekar_changing_2019, kizilay_incidence_2025}. This shifting epidemiology complicates screening, necessitating robust diagnostic tools which are capable of handling diverse clinical presentations and demographic variables \cite{wang_global_2024}.

At the same time, the medical definition of ROP has also changed. According to the new guidelines of \textit{International Classification of Retinopathy of Prematurity, Third Edition} (ICROP3), Plus Disease, known as the characteristic of severe disease in ROP, changes from a binary point to a continuous scale of vascular anomaly\cite{chiang_international_2021}. This revision acknowledges that there are pathological changes in a continuous form that are frequently ignored in binary classification systems. As a result, there can be considerable subjectivity in diagnosis with remarkably low consistency in diagnosing `Pre-Plus' disease among experts\cite{bolon-canedo_dealing_2015, cole_variability_2022}. This ambiguity requires systems robust enough to differentiate urgent vascular pathology from intermediate stages. 

Deep Learning has been an emerging solution for this screening problem. The foundational systems, DeepROP and i-ROP, achieved expert-level performance using Convolutional Neural Networks (CNNs) \cite{wang_automated_2018, redd_evaluation_2019}. There exists, however, a significant data divide problem. Although previous systems were based on private datasets of over 20,000 images, publicly available datasets, such as the Ostrava ROP dataset, involve small cohorts (N=188) that are highly imbalanced for the class distributions\cite{timkovic_retinal_2024}. This limitation makes the use of conventional architectures inefficient due to overfitting, which fails to identify the subtle vascular morphology that defines the ICROP3 spectrum.

Furthermore, the current state of the art (SOTA) models exhibits task fragmentation, often predicting either the broad stages of diagnosis or the plus disease separately, and in most cases, serve as `black box' that ignore the clinical priors. In the context of neonatal analysis, Gestational Age (GA), Post-conceptual Age (PA), and Birth Weight (BW) are known to be important priors that impact the level of suspicion. However, the current models use passive fusion, wherein the demographic features and the image features are concatenated only in the last layer. We argue that this late-fusion approach is suboptimal as it fails to leverage metadata to guide visual feature extraction. To overcome the above challenges, this research work proposes a \textbf{Context-Aware Asymmetric Ensemble Framework} (CAA Ensemble) that follows the Biomimetic AI paradigm. This study makes the following contributions by segregating Structural analysis and Vascular analysis for the ROP:

\begin{itemize}
    \item \textbf{Active Query Mechanism (MS-AQNet):} We propose a `Structure Specialist' that uses clinical metadata as dynamic query vectors to spatially gate visual feature extraction, addressing the drawbacks inherent with passive fusion.
    \item \textbf{Anatomy-Aware MIL (VascuMIL):} We introduce a ‘Texture Specialist’ that analyzes vascular topology maps (VMAP) using Multiple Instance Learning to detect Plus Disease.
    \item \textbf{Unified Multi-Task Synergy:} In contrast to the isolated classifiers, our synergistic fusion module combines orthogonal structural and vascular information to provide a comprehensive diagnostic profile, resolving conflicting information between the structural and vascular modalities in staging.
    \item \textbf{Data Efficiency \& Explainability:} The framework tackles the `Data Divide' problem, achieving SOTA results on a small public benchmark dataset, alongside providing transparent dual-stream clinical explainability.
\end{itemize}

% --- TABLE PLACED HERE TO APPEAR ON TOP OF NEXT PAGE ---
\begin{table*}[!t]
\centering
\caption{Comparative Analysis of Representative Methods in Automated Retinal Screening}
\label{tab:comparison}
\renewcommand{\arraystretch}{1.25} % Breathing room
\setlength{\tabcolsep}{4pt}        % Tighter horizontal padding to fit text

% TABULARX: The 'X' column auto-calculates width to fill the page exactly.
% We use 'l' (left) for fixed short columns, and 'X' for long text.
\begin{tabularx}{\textwidth}{@{}p{0.11\textwidth} p{0.11\textwidth} p{0.14\textwidth} p{0.09\textwidth} p{0.15\textwidth} X@{}}
\toprule
\textbf{Study} & \textbf{Primary Task} & \textbf{Method} & \textbf{Fusion \newline(Strategy)} & \textbf{Dataset \newline (Size \& Type)} & \textbf{Key Limitation} \\ 
\midrule

Wang \textit{et al.} \cite{wang_automated_2018} \newline (2018) & ROP Screening & DeepROP \newline (Inception-BN) & Image Only & Private \newline ($\sim$26k images) & Relies on ``brute force'' scale; overfits on small data. \\

Redd \textit{et al.} \cite{redd_evaluation_2019} \newline (2019) & ROP Plus & i-ROP \newline (U-Net + Inception) & Image Only & Private \newline ($\sim$4.9k images) & ``Black Box'' model; ignores clinical metadata. \\

Kang \textit{et al.} \cite{kang_multimodal_2021} \newline (2021) & Retinal Vascular & EfficientNet & Passive (Late) & Private \newline ($\sim$35k images) & Passive concatenation fails to guide feature extraction. \\

Li \textit{et al.} \cite{li_transfer_2023} \newline (2023) & Glaucoma & GMNNnet \newline (Multi-branch) & Active (Hybrid) & GM367 \newline ($N=367$ patients) & Good active fusion, but applied to Glaucoma, not ROP. \\

Zhou \textit{et al.} \cite{zhou_deep_2018} \newline (2018) & Diabetic Retinopathy & Deep MIL & Image Only & Kaggle DR \newline ($N=35,126$ images) & Standard MIL relies on texture; lacks vascular topology. \\

Sharafi \textit{et al.} \cite{sharafi_automated_2024} \newline (2024) & ROP Plus & SVM + Vessel Seg. & Image Only & Private \newline ($N=76$ images) & Relies on handcrafted features; not an end-to-end learning model. \\

Yenice \textit{et al.} \cite{kiran_yenice_automated_2024} \newline (2024) & ROP Detection & RegNetY002 & Image Only & Private \newline ($N=317$ infants) & Efficient architecture, but limited by small sample size. \\

Matten \textit{et al.} \cite{matten_multiple_2023} \newline (2023) & DR (OCTA) & MIL-ResNet14 & Image Only & Private \newline ($N=352$ images) & Effective MIL application, but lacks metadata integration. \\

Peng \textit{et al.} \cite{peng_mafe-net_2024} \newline (2024) & Retinal Seg. & MAFE-Net \newline (Ensemble) & Image Only & Public \newline ($N=88$ images) & Proves ensemble efficacy, but focuses on segmentation, not diagnosis. \\

Daho \textit{et al.} \cite{daho_improved_2023} \newline (2023) & DR Severity & Deep Fusion \newline (Mixup) & Active (Mixup) & Private \newline ($N=875$ images) & Advanced fusion (Mixup), but requires 3D OCTA data. \\

Deepthi \textit{et al.} \cite{deepthi_automated_2025} \newline (2025) & ROP Plus & Transformer \newline (Unsupervised) & Image Only & Ostrava ROP \newline Dataset & Uses unsupervised learning; misses clinical metadata. \\

Subramaniam \cite{subramaniam_image_2023} \newline (2023) & ROP Plus & GoogLeNet & Image Only & SP-ROP \newline ($N=440$ images) & Uses cost-sensitive loss, but lacks multi-modal integration. \\ 
\bottomrule
\end{tabularx}
\end{table*}

\section{Related Work}
\label{sec:related_work}
Automating ROP screening has progressed from manually crafted morphological filters to end-to-end deep learning models and more recently towards multi-modal attention-based models. This section reviews the literature across four key domains: ROP model evolution and the data divide, multi-modal fusion techniques, Multiple Instance Learning (MIL) for vascular topology, techniques for handling class imbalance. A brief comparison across these areas is given in Table \ref{tab:comparison}.

\subsection{CNN-based Deep Learning Models in ROP}
The application of CNNs to ROP detection was cemented by two landmark studies establishing the feasibility of automated screening. Wang \textit{et al.} \cite{wang_automated_2018} developed DeepROP, a two-stream network that comprised an Id-Net for identification and a Gr-Net for grading. With over 26,000 images used to train DeepROP, it achieved sensitivity of 96.6\%, eliminating manual feature engineering through deep neural networks. On the other hand, Redd \textit{et al.} \cite{redd_evaluation_2019} tested and confirmed the i-ROP DL system within ROP screening, employing a U-Net for image segmentation and an Inception-V1 model for classification to provide a quantitative vascular severity score on a 1 to 9 scale that had an AUC of 0.98 for Plus Disease, relating well to the spectrum indicated by ICROP3 within a continuous manner.
\cite{bai_multicenter_2023, wagner_development_2023}
However, recent external validations demonstrate the vulnerability of such heavy models in the presence of varied real-world data\cite{bai_multicenter_2023, wagner_development_2023}. Model performance tends to degrade if the populations or imaging devices are dissimilar from those in the training environment. This limitation is the key to the data divide problem. Ostrava ROP Dataset\cite{timkovic_retinal_2024} demonstrates that publicly available data are often restricted to limited cohorts ($N \approx. 200$) with high class imbalance. Reviews by Nakayama \textit{et al.} \cite{nakayama_fairness_2023} illustrate that generic transfer learning models generally suffer from overfitting on small datasets. Yenice \textit{et al.} \cite{kiran_yenice_automated_2024} attempted to address this by using lighter models such as MobileNet on small datasets ($N=320$), but standard CNN models still lack the inductive biases to learn complex vascular morphology from small amounts of data.

\subsection{Multi-Modal Fusion Models}
The integration of fundus images with clinical information is critical to imitating a holistic decision-making process by a clinician. However, the majority of fusion literature is based on passive fusion based on late feature concatenation. For instance, Kang \textit{et al.} \cite{kang_multimodal_2021} developed a multimodal platform that integrated fundus photo, OCT scans, and angiography to diagnose Retinal Vascular Disease. Their model succeeded, but it was merely doing feature flattening, followed by feature concatenation at the last layer. According to El-Ateif and Idri\cite{el-ateif_multimodality_2024}, this late fusion is inefficient since it does not enable one modality to guide another during feature extraction.

We argue that this is suboptimal for ROP, where clinical context should actively guide visual search. Emerging literature supports this hypothesis; for example, in glaucoma detection, Li \textit{et al.} \cite{li_transfer_2023} proposed GMNNnet based on multi-branch networks for analyzing meta-data in conjunction with images. For the task of DR, more advanced approaches for fusions such as Multi-Scale Attention\cite{al-antary_multi-scale_2021} or Manifold Mixup\cite{daho_improved_2023}, demonstrate that it is beneficial to work with deep feature spaces. Cai \textit{et al.} \cite{cai_uni4eye_2022} proposed Uni4Eye based on the unified embedding for simultaneously processing 2D and 3D modalities.

\subsection{Vascular Topology and Multiple Instance Learning (MIL)}
While pixel-level segmentation works for macular structures, finding Plus disease is a needle-in-a-haystack problem. Pathological tortuosity is likely to be a patchwork process, and global pooling may reduce these patchwork characteristics. The MIL model considers images as bags of patches as pioneered in Diabetic Retinopathy \cite{zhou_deep_2018} and widefield OCTA \cite{matten_multiple_2023}. However, traditional MIL usually relies on generic texture descriptors that are confounded by retinal pigmentation. We argue that ROP cases require specific anatomical descriptors. Quantitative studies support this: for instance, curvature measures are more accurate as a severity predictor than vessel width, as found by Turior \textit{et al.} \cite{turior_quantification_2013}. In a recent study, Huang \textit{et al.} \cite{huang_computer-aided_2024} demonstrated a significant relationship  $(\text{P} < 0.0001)$ using polynomial curve fitting between tortuosity measures and stages of ROP. Similarly, Sharafi \textit{et al.} \cite{sharafi_automated_2024} achieved high diagnostic accuracy using B-COSFIRE filters for vessel segmentation, proving morphological features are robust disease proxies. 

\subsection{Methods Addressing Class Imbalance and Small Data}
Medical datasets are inherently imbalanced \cite{wang_global_2024}. The usual technique used to reduce this problem is GAN-based data augmentation, but that would result in intra-class mode collapse where models fail to generate diverse minority samples \cite{ding_leveraging_2024}. To address this problem, Deepthi \textit{et al.} \cite{deepthi_automated_2025} introduced unsupervised curriculum learning-based pseudo-labeling, and Subramaniam \textit{et al.} \cite{subramaniam_image_2023} proposed image harmonization and cost-sensitive loss. However, these data-centric methods do not solve the fundamental architectural misalignment of treating structure and texture as a single task.  Recent research by Peng \textit{et al.} \cite{peng_mafe-net_2024} implies that ensemble approaches, involving multiple models, provide superior results, supporting our asymmetric approach to handling varying feature scales \cite{raja_sankari_automated_2023}.

% --- FIGURE 1: FRAMEWORK (Double Column) ---
\begin{figure*}[!t]
\centering
\includegraphics[width=0.95\linewidth]{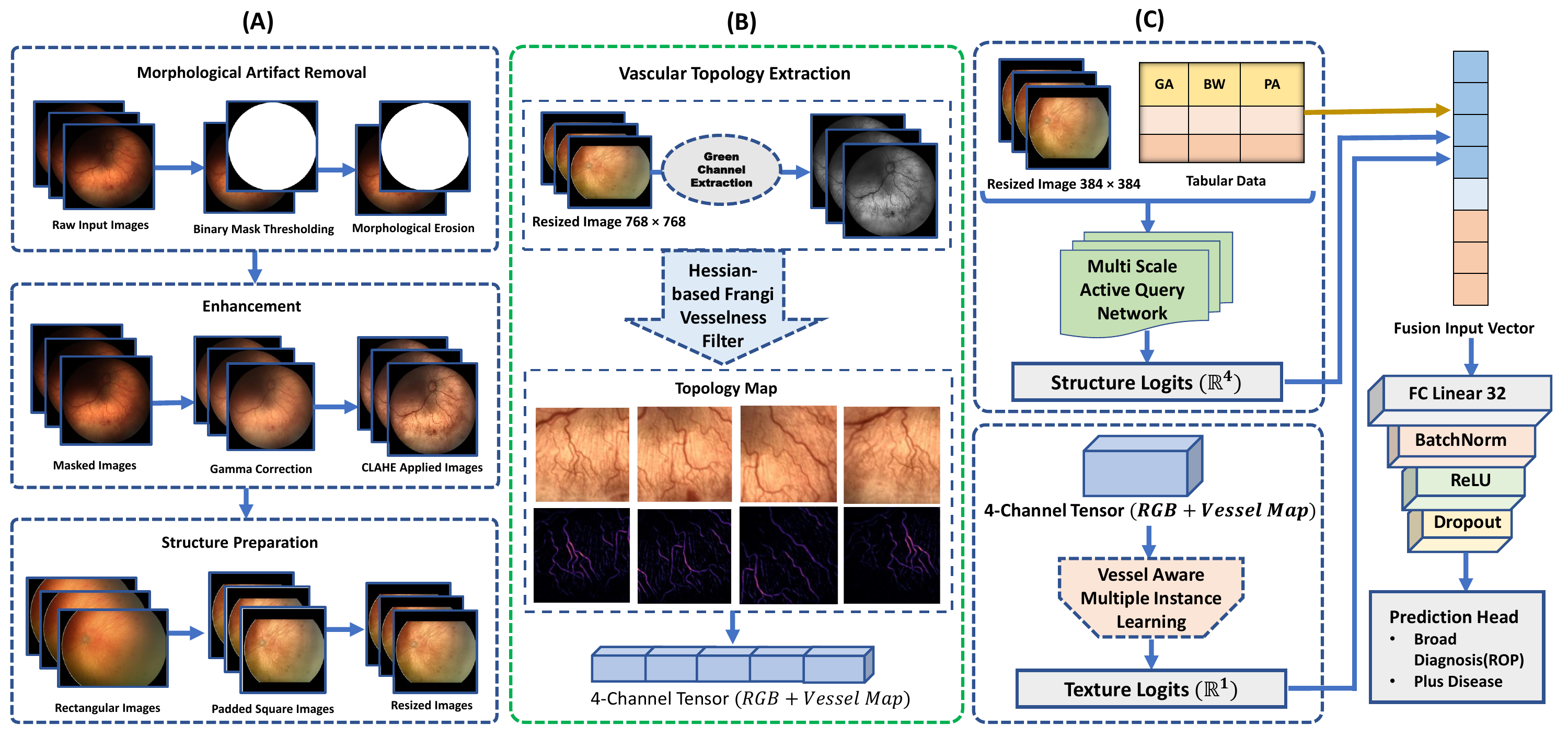}
\caption{Overview of the Context-Aware Asymmetric Ensemble. \textbf{(A) Intelligent Data Engineering:} Morphological artifact removal and geometry-preserving resizing ($384 \times 384$) for structural analysis. \textbf{(B) Vascular Topology Extraction:} Generation of VMAPs from high-resolution inputs ($768 \times 768$) to form 4-channel tensors for textural analysis. \textbf{(C) Processing \& Fusion:} Parallel execution of the Structure Specialist (MS-AQNet) and Texture Specialist (VascuMIL), followed by synergistic fusion of logits with re-injected clinical metadata.}
\label{fig:framework}
\end{figure*}
% -------------------------------------------

\section{Methodology}
\label{sec:methodology}
\subsection{ Proposed Context-Aware Asymmetric Ensemble (CAA Ensemble)}
\label{subsec:framework}
Retinopathy of Prematurity (ROP) manifests itself in two different ways: macro irregularities, as large-scale structural anomalies, such as ridges and detachments, and as fine-grained micro-vascular irregularities, like tortuosity~\cite{icrop_revisited_2005}. To model these aspects, we introduce the Context-Aware Asymmetric Ensemble (CAA Ensemble). The subsequent sections detail the intelligent data engineering strategy, the specialized architectural streams, and the synergistic fusion mechanism that comprise this framework.

\subsection{Intelligent Data Engineering}
The proposed framework follows a resolution bifurcation approach (Fig. \ref{fig:framework} Panel B and C). The traditional resize operation removes much of the detail necessary for a Plus Disease identification, while a method that preserves resolution may introduce noise that makes it difficult for the structural classification. To satisfactorily address both requirements, a two-path approach is utilized:

\begin{itemize}
    \item \textbf{Structure Stream:} This stream focuses on disease staging. The stream uses global images of the fundus at a uniform resolution of $384 \times 384$ pixels and is combined with clinical metadata through the Multi-Scale Active Query Network (MS-AQNet).
    \item \textbf{Texture Stream:} It is designed to target Plus Disease. It analyzes high-resolution maps of the vascular topology ($768 \times 768$) through the proposed VascuMIL model.
\end{itemize}

\subsubsection{Morphological Artifact Removal}
To avoid the incorporation of spurious correlations due to aperture artifacts in the model, a morphological cleaning operation is used. First, a binary thresholding operation is performed to segregate the background from the foreground in the fundus image. We then apply morphological erosion with a $3\times3$ kernel to shrink the mask, which gets rid of the glowing edges. Finally, gamma correction with $\gamma=1.5$ and CLAHE (Contrast Limited Adaptive Histogram Equalization) is also applied on the masked images (Fig. \ref{fig:framework} Panel A).

\subsubsection{Vascular Topology Extraction (For Texture Stream)}
To provide explicit geometric priors, we generate a \textit{Vascular Topology Map (VMAP)}. Given the green channel $\mathbf{I}_G$ which is chosen for maximal vascular contrast, we compute the Frangi vesselness $\mathcal{V}_f(\mathbf{x})$ based on the sorted eigenvalues ($|\lambda_1| \le |\lambda_2|$) of the Hessian matrix at multiple scales \cite{frangi1998multiscale}. The vesselness probability is computed as:

\begin{equation}
\mathcal{V}_f(\mathbf{x}) = \begin{cases} 
0 & \text{if } \lambda_2 > 0 \\
\exp\left(-\frac{\mathcal{R}_{\mathcal{B}}^2}{2\beta^2}\right) \left(1 - \exp\left(-\frac{\mathcal{S}^2}{2c^2}\right)\right) & \text{otherwise}
\end{cases}
\label{eq:vesselness}
\end{equation}

where $\mathcal{R}_{\mathcal{B}} = |\lambda_1|/|\lambda_2|$ serves as the \textit{blobness measure} to distinguish tubular structures from background blobs, and $\mathcal{S} = \sqrt{\lambda_1^2 + \lambda_2^2}$ measures \textit{structureness} to suppress noise. The constants $\beta$ and $c$ tune the sensitivity to these respective measures. The sparse map is then stacked depth-wise on the RGB patches to construct input tensors of 4-channels, forcing the network to correlate the color texture to the geometric structure of the vasculature.

% --- FIGURE 2: MS-AQNET (Double Column) ---
\begin{figure*}[!t]
\centering
\includegraphics[width=0.95\linewidth]{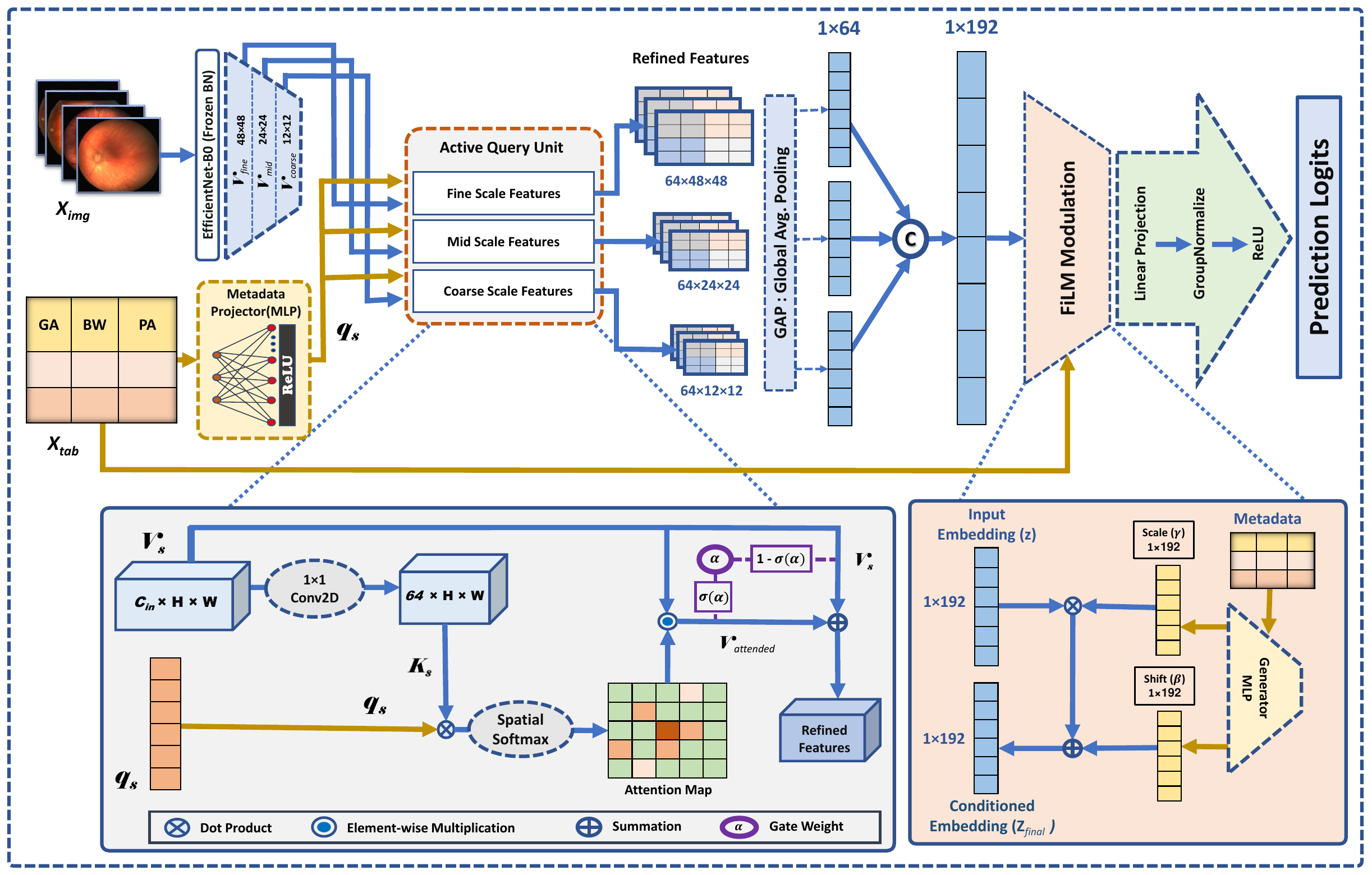}
\caption{Architecture of the Multi-Scale Active Query Network (MS-AQNet). \textbf{(A)} The hierarchical pipeline extracting features at three spatial scales via an EfficientNet-B0 backbone. \textbf{(B)} Detail of the \textbf{Active Query Unit}, showing how the projected clinical query ($\mathbf{q}_s$) generates a spatial attention map via dot-product similarity, regulated by a learnable gate ($\alpha$). \textbf{(C)} Detail of the \textbf{FiLM Block}, illustrating the global affine transformation ($\gamma, \beta$) of the visual embedding based on clinical priors.}
\label{fig:msaqnet}
\end{figure*}
% ------------------------------------------

\subsection{Proposed Multi-Scale Active Query Network (MS-AQNet)}
To determine structural biomarkers such as demarcation lines and ridges, we introduce the \textit{Multi-Scale Active Query Network} (MS-AQNet) depicted in Fig. \ref{fig:msaqnet}. In contrary to conventional fusion approaches that follow a passive fusion process, MS-AQNet follows an active querying process where clinical priors (Gestational Age, Birth Weight) act as an active query signal instead. This introduces a conditional inductive bias where the physiological risk profile of the individual queries the feature maps. The subsequent subsections detail the architectural implementation of this approach, covering stable feature extraction, the localized spatial attention mechanism, and global semantic calibration via modulation.

\subsubsection{Hierarchical Feature Extraction and Stability}
We use an EfficientNetB0 backbone\cite{tan2020efficientnetrethinkingmodelscaling} to produce hierarchical maps $\mathbf{V} = \{ \mathbf{V}_s \}_{s=1}^{3}$ from the input $\mathbf{X}_{img}$. This multi-scale approach captures both minute vascular abnormalities and global retinal haze.

Considering the small data set ($N=188$), the batch size during training is restricted ($B=16$). To reduce the problem of internal covariate shift, we apply \textit{Frozen Batch Normalization}. During the training phase, the mean and variance statistics ($\mu_{BN}, \sigma^2_{BN}$) of the global statistics are frozen to the pre-trained ImageNet statistics to decouple feature extraction stability from small-batch stochasticity.

\subsubsection{The Active Query Mechanism}
The core \textit{Active Query Unit} introduces the clinical context, a normalized vector $\mathbf{x}_{tab}$, into the spatial domain.

\textit{Latent Query Projection:} A \textit{Metadata Projector Network (MLP)} bridges the modality gap, mapping scalars into a high-dimensional latent query vector $\mathbf{q}_s$ as defined in \eqref{eq:query_proj}:

\begin{equation}
\mathbf{q}_s = \Phi_s(\mathbf{x}_{tab}) = \text{ReLU}[\mathbf{W}_{2,s} \cdot \text{ReLU}(\mathbf{W}_{1,s} \cdot \mathbf{x}_{tab})] \in \mathbb{R}^{d_{vis}}
\label{eq:query_proj}
\end{equation}

where $d_{vis}=64$ represents the standardized channel depth and $\mathbf{W}_{.,s}$ represents the learnable weight matrices.

\textit{Spatial Dot-Product Attention:}
The visual feature map $\mathbf{V}_s$ is projected into a key tensor $\mathbf{K}_s$. We compute a spatial attention map $\mathbf{A}_s$ by the dot-product between the clinical query $\mathbf{q}_s$ and visual features\cite{vaswani2023attentionneed} which is calculated as \eqref{eq:dot_product}:

\begin{equation}
e_{ij}^s = \frac{\mathbf{q}_s^\top \mathbf{k}_{ij}^s}{\sqrt{d_{vis}}}
\label{eq:dot_product}
\end{equation}

where $\mathbf{k}_{ij}^s$ is the feature vector at spatial position $(i,j)$. A spatial softmax converts these scores into a probability distribution, given by \eqref{eq:spatial_softmax}:

\begin{equation}
\mathbf{A}_{ij}^s = \frac{\exp(e_{ij}^s)}{\sum_{u=1}^{H}\sum_{v=1}^{W} \exp(e_{uv}^s)}
\label{eq:spatial_softmax}
\end{equation}

where $H, W$ are the spatial dimensions, and high magnitude in $\mathbf{A}_s$ indicates that the regions are structurally relevant given the patient's risk profile.

\textit{Adaptive Residual Gating:}
To prevent over-reliance on metadata, a learnable gate $\alpha_s$ controls the mixing ratio as described in \eqref{eq:adaptive_gating}:

\begin{align}
g_s &= \sigma(\alpha_s) \nonumber \\
\tilde{\mathbf{V}}_s &= (1 - g_s) \cdot \mathbf{V}_s + g_s \cdot (\mathbf{A}_s \odot \mathbf{V}_s)
\label{eq:adaptive_gating}
\end{align}

where $\sigma$ is the sigmoid function, $\odot$ denotes element-wise multiplication, and we initialize $\alpha_s = -2.0$ ($g_s \approx 0.12$), causing the model initially to become biased towards the visual features.

\subsubsection{Global Aggregation and FiLM Modulation}
Following the active query, Global Average Pooling (GAP) is applied to compress the refined feature into a vector $\mathbf{z}$. We employ \textit{Feature-wise Linear Modulation (FiLM)}\cite{perez2017filmvisualreasoninggeneral} for global semantic conditioning. Unlike standard FiLM implementations that use recurrent networks for linguistic input, we define the generator $\Psi$ as a \textit{Multi-Layer Perceptron (MLP)} to process the static clinical vector. This generator predicts scale ($\boldsymbol{\gamma}$) and shift ($\boldsymbol{\beta}$) modulation vectors from $\mathbf{x}_{tab}$ via \eqref{eq:film_params}:

\begin{equation}
\boldsymbol{\gamma}, \boldsymbol{\beta} = \Psi_{MLP}(\mathbf{x}_{tab}) \in \mathbb{R}^{192}
\label{eq:film_params}
\end{equation}
\begin{equation}
\mathbf{z}_{final} = \boldsymbol{\gamma} \odot \mathbf{z} + \boldsymbol{\beta}
\label{eq:film_modulation}
\end{equation}

This affine transformation adjusts the feature distribution dynamically, effectively calibrating the decision boundary based on the degree of physiological severity.

\subsubsection{Robust Classification Head}
We use \textit{Group Normalization (GN)}\cite{wu2018groupnormalization} in the classification head to ensure stability in smaller batch sizes. The GN calculates the statistics on the groups of channels in each sample, accounting for the estimation noise in studies involving small cohorts. The output is a four-class distribution based on structural severity.

% --- FIGURE 3: VASCUMIL (Single Column) ---
\begin{figure}[!t]
\centering
\includegraphics[width=\linewidth]{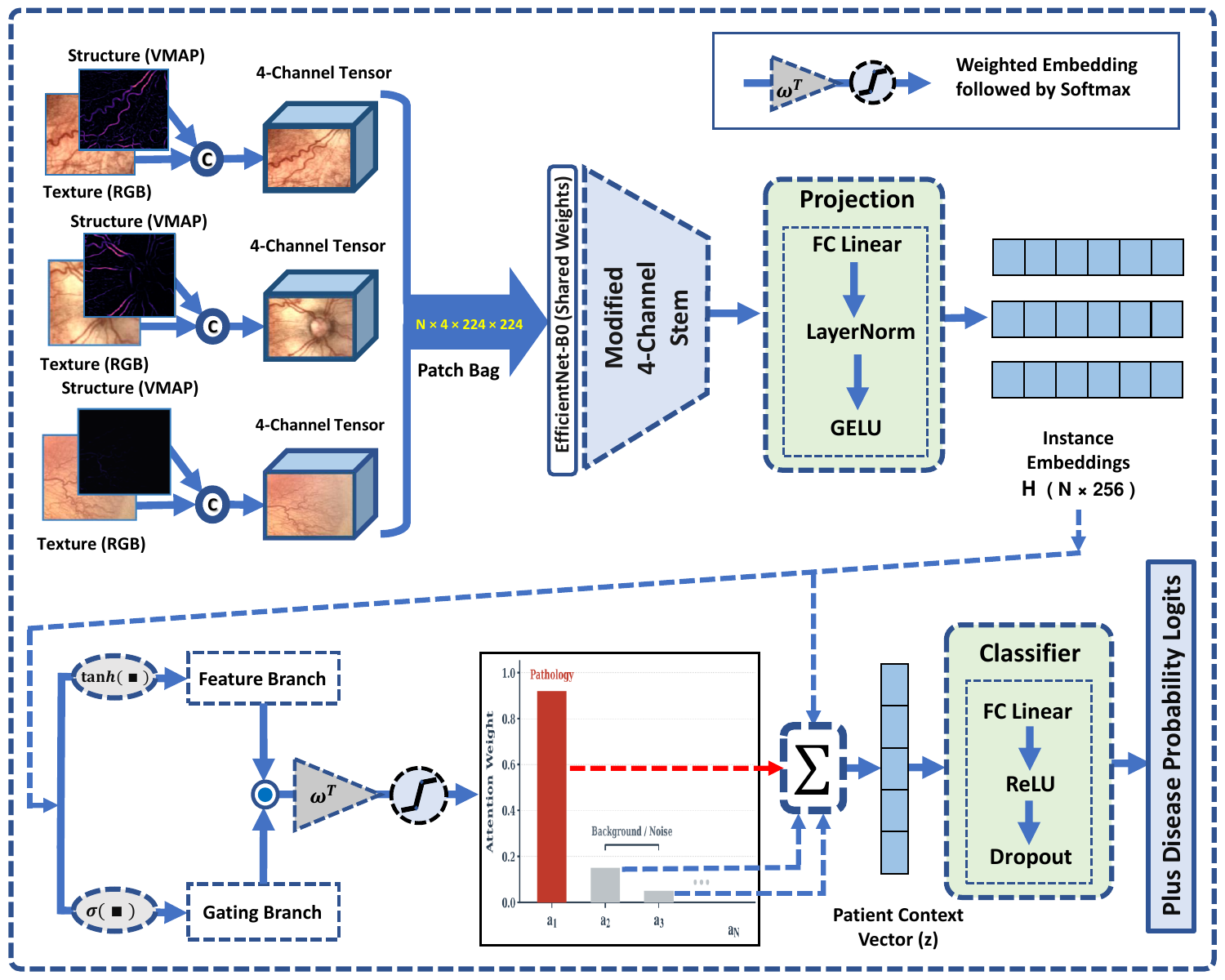}
\caption{\textbf{Architecture of the VascuMIL Network.} \textbf{(Top)} \textit{Feature Extraction:} 4-channel input tensors (RGB + Vascular Topology Maps) are encoded via a weight-shared backbone with a modified stem. \textbf{(Bottom)} \textit{Gated Attention:} A learnable gating mechanism ($\tanh \odot \sigma$) assigns importance weights to instances, isolating pathological signals (Red) from background noise (Grey) to aggregate the final patient context vector $\mathbf{z}$.}
\label{fig:vascumil}
\end{figure}
% ------------------------------------------

\subsection{Proposed Vascular-Aware Multiple Instance Learning Network (VascuMIL)}
\label{subsec:vascumil}

While MS-AQNet captures global structural pathologies, diagnosing \textit{Plus Disease} relies on identifying micro-vascular abnormalities (tortuosity, dilation) often lost in downsampled images. To address this, we propose \textit{VascuMIL}, as shown in Fig. \ref{fig:vascumil}, a texture-specialist formulated within a Multiple Instance Learning (MIL) framework~\cite{quellec_multiple_2017}. VascuMIL aggregates evidence from high-resolution patches, enabling localized vascular pathology detection while maintaining patient-level decision integrity. The following subsections detail this texture-driven pipeline, covering the four-channel vascular enhancement, gated instance encoding, and the final diagnostic aggregation process.

\subsubsection{4-Channel Vascular Enhancement}
In order to incorporate a strong inductive bias towards the blood vessel topological information, a \textit{Vessel Map (VMAP)} is calculated based on Frangi vesselness filtering (detailed in Section II-B). Given a high-resolution patch $\mathbf{I}_{RGB}$, a vessel probability map, $\mathbf{M}_{vess}$, is calculated, and then stacked along the depth dimension to create an advanced four-channel feature tensor $\mathbf{x}_k$ via \eqref{eq:4channel}:

\begin{equation}
\mathbf{x}_k = \text{Concat}(\mathbf{I}_{RGB}, \mathbf{M}_{vess}) \in \mathbb{R}^{4 \times H \times W}
\label{eq:4channel}
\end{equation}

This formulation forces the feature extractor to learn simultaneously from chromatic texture (RGB) and structural topology (VMAP).

\subsubsection{Bag Construction and Instance Encoding}
Patient examinations are represented as a bag of data $\mathbf{X} = \{\mathbf{x}_1, \dots, \mathbf{x}_N\}$ with $N$ patches. An \textit{EfficientNet-B0} backbone network, denoted as $f_{cnn}(\cdot; \theta_{enc})$, with a modified input stem is used where the weights of the first convolutional layer are modified to be of four channels. The weights for the extra channels are initialized using Kaiming initialization\cite{kaiming_he2015}. The pre-trained weights for the RGB channels are used to adapt quickly.

Each patch $\mathbf{x}_k$ is mapped to a low-dimensional embedding $\mathbf{h}_k$ as given in \eqref{eq:instance_emb}:

\begin{equation}
\mathbf{h}_k = \phi[f_{cnn}(\mathbf{x}_k; \theta_{enc})] \in \mathbb{R}^{L}
\label{eq:instance_emb}
\end{equation}

where $\phi$ is a projection head mapping features to a compact space of dimension $L=256$.

\subsubsection{Gated Attention Mechanism}
Standard pooling risks diluting sparse pathological signals. We employ a \textit{Gated Attention Mechanism}\cite{ilse2018attentionbaseddeepmultipleinstance} to assign learnable importance scores $a_k$ using \eqref{eq:gated_attention}:

\begin{equation}
a_k = \frac{\exp\left[ \mathbf{w}^\top \left\{ \tanh(\mathbf{V}\mathbf{h}_k) \odot \sigma(\mathbf{U}\mathbf{h}_k) \right\} \right]}{\sum_{j=1}^{N} \exp\left[ \mathbf{w}^\top \left\{ \tanh(\mathbf{V}\mathbf{h}_j) \odot \sigma(\mathbf{U}\mathbf{h}_j) \right\} \right]}
\label{eq:gated_attention}
\end{equation}

where $\mathbf{w} \in \mathbb{R}^{M \times 1}$, $\mathbf{V} \in \mathbb{R}^{M \times L}$, and $\mathbf{U} \in \mathbb{R}^{M \times L}$ are learnable weights with attention dimension $M=128$, and $\odot$ denotes element-wise multiplication. In this architecture, the hyperbolic tangent branch, $\tanh(\cdot)$, serves as a feature extractor, while the sigmoid branch, $\sigma(\cdot)$, functions as a soft gating mechanism. The dual-branch structure adjusts activations of these features by down-weighting non-informative background patches and emphasizing those showing high tortuosity values.

\subsubsection{Aggregation and Classification}
The final patient-level representation $\mathbf{z}$ is the attention-weighted sum of instance embeddings, defined as \eqref{eq:aggregation}:

\begin{equation}
\mathbf{z} = \sum_{k=1}^{N} a_k \mathbf{h}_k
\label{eq:aggregation}
\end{equation}

This context vector feeds into a classification head $g_{cls}$ which is parameterized by learnable weights $\theta_{cls}$ and structured as an MLP ($d_{hid}=64$, Dropout $p=0.2$) to predict the probability of Plus Disease as \eqref{eq:plus_pred}:

\begin{equation}
\hat{y}_{plus} = \sigma(g_{cls}(\mathbf{z}; \theta_{cls}))
\label{eq:plus_pred}
\end{equation}

This architecture ensures diagnosis is driven by the most suspicious vascular features, mimicking expert search-and-confirm cognition.

\subsection{Synergistic Feature Fusion}

To integrate the severity (Stage) with vascular activity (Plus) dimensions, a \textit{Fusion Meta-Learner} is used. A unified vector $\mathbf{z}_{fusion}$ is generated through the concatenation process involving the un-normalized logit vector from MS-AQNet ($\mathbf{L}_{struct} \in \mathbb{R}^4$) and the logit scalar from VascuMIL ($L_{tex} \in \mathbb{R}^1$), with reinserted clinical metadata $\mathbf{x}_{tab} \in \mathbb{R}^3$, as shown in \eqref{eq:fusion_vector}:

\begin{equation}
\mathbf{z}_{fusion} = \text{Concat}(\mathbf{L}_{struct}, L_{tex}, \mathbf{x}_{tab}) \in \mathbb{R}^{8}
\label{eq:fusion_vector}
\end{equation}

The reintroduction of metadata allows stream confidence to be calibrated for patient demographics. The fusion vector is then passed to a shallow MLP ($f_{mlp}$) where ($d_{in}=8, d_{out}=32$), producing a hidden representation $\mathbf{h}_{fusion}$ which splits into two task-specific heads characterized by \eqref{eq:head_diag} and \eqref{eq:head_plus}:

\begin{align}
\hat{Y}_{diag} &= \text{Softmax}(\mathbf{W}_{diag} \cdot \mathbf{h}_{fusion} + \mathbf{b}_{diag}) \label{eq:head_diag} \\
\hat{y}_{plus} &= \sigma(\mathbf{w}_{plus}^\top \mathbf{h}_{fusion} + b_{plus}) \label{eq:head_plus}
\end{align}

where $\mathbf{W}_{diag}, \mathbf{w}_{plus}$ denote the learnable weights and $\mathbf{b}_{diag}, b_{plus}$ denote the bias terms. This allows for dynamic prioritization of Texture Stream content for vascular signals (e.g. Aggressive ROP) and stage-defined disease to be determined from Structure Stream content.

\subsection{Optimization and Loss Functions}
\label{subsec:optimization}

To ensure convergence on this imbalanced dataset, we employ specific loss strategies tailored to the learning dynamics of each stream.

\subsubsection{Structural Optimization with Deep Supervision}
MS-AQNet uses \textit{Deep Supervision} to overcome the problem of gradient vanishing. Auxiliary classifiers are connected to the first two Active Query Units. The overall structural loss $\mathcal{L}_{struct}$ combines the primary prediction $\hat{y}_{main}$ and extra heads $\hat{y}_{aux,k}$ against ground truth $y$ based on \eqref{eq:loss_struct}:

\begin{equation}
\mathcal{L}_{struct} = \mathcal{L}_{focal}(\hat{y}_{main}, y) + \lambda \sum_{k=1}^{2} \mathcal{L}_{focal}(\hat{y}_{aux,k}, y)
\label{eq:loss_struct}
\end{equation}

wherein $\lambda=0.2$. To address the severe class imbalance, a class-weighted variant of the Focal Loss function\cite{lin2018focallossdenseobject} is applied. A clinical prioritization weight set ($w_{severe}=5.0$, $w_{normal}=0.5$) is assigned to the balancing factor $\alpha_t$, giving a large penalty to the false negatives in sight-threatening conditions. The loss function is given by \eqref{eq:focal}:

\begin{equation}
\mathcal{L}_{focal}(p_t) = -\alpha_t (1 - p_t)^\gamma \log(p_t)
\label{eq:focal}
\end{equation}

where $\gamma=2.0$ is the focusing parameter. This reshapes the gradient landscape to down-weight easy negatives, forcing the model to focus on hard, minority examples.

\subsubsection{Vascular Optimization via Weighted BCE}
The VascuMIL is optimized through \textit{Weighted Binary Cross-Entropy (BCE)}. Given the fact that positive bags are rare (with a probability of $\approx 3\%$), standard objectives introduce bias in decision boundaries. For this purpose, a weighted loss is considered for bag label $y$ and prediction $\hat{y}$ which is calculated by \eqref{eq:loss_tex}:

\begin{equation}
\mathcal{L}_{tex} = - [w_{pos} \cdot y \log(\sigma(\hat{y})) + (1-y) \log(1-\sigma(\hat{y}))]
\label{eq:loss_tex}
\end{equation}

where $w_{pos}$ scales gradients based on inverse class frequency to ensure that rare positive instances provide a sufficient learning signal.

\subsubsection{Joint Multi-Task Fusion Objective}
The Fusion Meta-Learner seeks to optimize a \textit{joint multi-task objective}, given by equation \eqref{eq:loss_total}, which combines both the multi-class diagnosis predictions $\hat{Y}_{diag}$ and the binary Plus Disease prediction $\hat{y}_{plus}$:

\begin{equation}
\mathcal{L}_{total} = \mathcal{L}_{CE}(\hat{Y}_{diag}, Y_{diag}) + \mathcal{L}_{BCE}(\hat{y}_{plus}, y_{plus})
\label{eq:loss_total}
\end{equation}

\section{Experimental Setup}
\label{sec:experimental}

\begin{table*}[t!]
\centering
\caption{Demographic and Clinical Characteristics of the Dataset across Training, Validation, and Test Splits}
\label{tab:demographics}
\renewcommand{\arraystretch}{1.25} 
\setlength{\tabcolsep}{0pt} % Let latex calculate the perfect spacing

% \textwidth forces the table to fit the page exactly
% @{\extracolsep{\fill}} spreads the columns out evenly
\begin{tabular*}{\textwidth}{@{\extracolsep{\fill}} l c c c c c}
\toprule
\textbf{Characteristic} & \textbf{Total} & \textbf{Training} & \textbf{Validation} & \textbf{Test Set} & \textbf{\textit{p}-value} \\
& (N=188) & (Fold 0) & (Fold 0) & (Held-out) & (Tr vs Val) \\
\midrule

\multicolumn{6}{l}{\textit{\textbf{Sample Size}}} \\
\hspace{1em} Subjects (Patients) & 188 & 137 & 34 & 17 & - \\
\hspace{1em} Images ($n$) & 6004 & 4261 & 1075 & 668 & - \\
\midrule

\multicolumn{6}{l}{\textit{\textbf{Clinical Metrics (Mean $\pm$ SD)}}} \\
\hspace{1em} Gestational Age (wks) & 31.0 $\pm$ 5.5 & 31.1 $\pm$ 5.5 & 30.5 $\pm$ 4.7 & 31.4 $\pm$ 6.4 & $<0.001$ \\
\hspace{1em} Birth Weight (g) & 1674 $\pm$ 1032 & 1703 $\pm$ 1028 & 1543 $\pm$ 943 & 1700 $\pm$ 1170 & $<0.001$ \\
\hspace{1em} Post-concept. Age (wks) & 40.1 $\pm$ 9.1 & 40.8 $\pm$ 10.3 & 37.6 $\pm$ 3.8 & 39.2 $\pm$ 5.2 & $<0.001$ \\
\midrule

\multicolumn{6}{l}{\textit{\textbf{Broad Diagnosis Distribution}}} \\
\hspace{1em} Normal & 2980 (49.6\%) & 2149 (50.4\%) & 550 (51.2\%) & 281 (42.1\%) & - \\
\hspace{1em} Mild / Treated & 676 (11.3\%) & 342 (8.0\%) & 223 (20.7\%) & 111 (16.6\%) & - \\
\hspace{1em} \textbf{Severe ROP} & 1053 (17.5\%) & 832 (19.5\%) & 143 (13.3\%) & 78 (11.7\%) & - \\
\hspace{1em} Other Pathology & 1295 (21.6\%) & 938 (22.0\%) & 159 (14.8\%) & 198 (29.6\%) & - \\
\midrule

\multicolumn{6}{l}{\textit{\textbf{Vascular Status}}} \\
\hspace{1em} \textbf{Plus Disease} & 629 (10.5\%) & 470 (11.0\%) & 81 (7.5\%) & 78 (11.7\%) & - \\

\bottomrule
\end{tabular*}

% Footnote is now OUTSIDE the tabular, preventing column distortion
\vspace{1mm}
\parbox{\textwidth}{
    \centering \footnotesize 
    \textit{Note: \textit{p}-values calculated via independent t-test between Training and Validation splits. Significant differences ($p<0.001$) in clinical metrics arise from the Stratified Group K-Fold strategy, which prioritizes preventing data leakage over perfectly matching feature distributions in small cohorts ($N=188$).}
}
\end{table*}

\subsection{Dataset and Demographics}
In this study, Retinal Image Dataset of Infants and Retinopathy of Prematurity~\cite{timkovic_retinal_2024} was used. It comprises 6,004 longitudinal fundus images of 188 preterm infants. These images were taken with Clarity RetCam 3, Natus RetCam Envision, and Phoenix ICON cameras. The combination of diverse imaging devices and pixel resolutions of $640 \times 480$, $1440 \times 1080$ and $1240 \times 1240$ poses a tough benchmarking challenge in terms of normalization. The patients fall in the high-risk group with mean gestational age of $31.0 \pm 5.5$ weeks and mean birth weight of $1674 \pm 1032$ g. These characteristics, as shown in Table \ref{tab:demographics}, show a strong class imbalance problem common in screening data with minority classes of severe ROP and Plus Disease at 17.5\% and 10.5\%, respectively.

This split was carried out patient-wise with very clear segregation. One-tenth of the total patient population ($N=17$) was set aside for an unseen Test Set, which was stratified to maintain a balanced representation of samples across all classes. The other 90\% ($N = 171$) comprised the training dataset for the purpose of stratified 5-fold cross-validation analyses. A very strict criterion was adopted to ensure that all images belonging to a patient were assigned to a single fold, thus circumventing potential data leaks inherent in patient-specific vascular features. This method, while introducing significant variability ($p < 0.001$) in patient-level distributions across the different folds owing to the small population size, helps to ensure a fair estimate of actual generalization capabilities.

\subsection{Target Engineering and Clinical Taxonomy}

% --- TABLE: TAXONOMY (Single Column) ---
\begin{table}[t!]
\centering
\caption{Clinical Severity Remapping Schema}
\label{tab:taxonomy}
\renewcommand{\arraystretch}{1.2}
\resizebox{\columnwidth}{!}{% Forces table to fit single column width
\begin{tabular}{l l l}
\toprule
\textbf{Target Class} & \textbf{Codes (DG)} & \textbf{Clinical Rationale} \\ 
\midrule
\textbf{0: Normal} & 0 & Healthy \\ 
\textbf{1: Mild} & 1, 2, 9 & \textbf{Observe:} Low immediate risk. \\ 
\textbf{2: Severe} & \textbf{3}, 4-8 & \textbf{Refer:} Pre-threshold/Threshold ROP. \\ 
\textbf{3: Other} & 10-13 & \textbf{Exclude:} Non-ROP (e.g., Hemorrhage). \\ 
\bottomrule
\end{tabular}%
}
\vspace{-2mm} 
\end{table}
% ---------------------------------------

Given the long-tail imbalance, we aggregated the 13 granular codes into a 4-Class Clinical Severity Scale (Table \ref{tab:taxonomy}). An important consideration was to assign Stage 2 (DG3) to the Severe category, based on ETROP guidelines~\cite{good_final_2004}, which classify Stage 2 with Plus Disease as requiring intervention. Stage 2, along with mild disease, is prone to false negatives; therefore, the classifier had to favor high sensitivity to align with referable ROP benchmarks~\cite{redd_evaluation_2019, wang_automated_2018} to maximize patient safety.

\begin{table*}[!t]
\centering
\begin{threeparttable}
\caption{Comparative Diagnostic Performance on Test Set: Baseline vs. Specialists vs. Ensemble}
\label{tab:clinical_performance}
\renewcommand{\arraystretch}{1.25} 
\setlength{\tabcolsep}{0pt} 

\begin{tabular*}{\textwidth}{@{\extracolsep{\fill}} l l c c c c c c } 
\toprule
\textbf{Task / Class} & \textbf{Model} & \textbf{Sens.} & \textbf{Spec.} & \textbf{Prec.} & \textbf{F1} & \textbf{AUC} & \textbf{Kappa} \\ 
\midrule \midrule

\multicolumn{8}{c}{\textit{\textbf{A. Broad Diagnosis (4-Class Structure)}}} \\ 
\midrule
\multirow{3}{*}{\textbf{Normal}} 
 & Baseline CNN & 0.648 & 0.914 & 0.861 & 0.740 & 0.894 & - \\
 & MS-AQNet & 0.573 & 0.936 & 0.915 & 0.705 & 0.833 & - \\
 & \textbf{CAA Ensemble} & \textbf{0.954} & \textbf{0.963} & \textbf{0.969} & \textbf{0.962} & \textbf{0.994} & - \\ \midrule
 
\multirow{3}{*}{\textbf{Mild / Treated}} 
 & Baseline CNN & 0.316 & 0.927 & 0.389 & 0.349 & 0.811 & - \\
 & MS-AQNet & 0.214 & 0.878 & 0.267 & 0.237 & 0.784 & - \\
 & \textbf{CAA Ensemble} & \textbf{0.971} & \textbf{0.968} & \textbf{0.862} & \textbf{0.913} & \textbf{0.996} & - \\ \midrule
 
\multirow{3}{*}{\textbf{Severe ROP}} 
 & Baseline CNN & 0.935 & 0.866 & 0.509 & 0.660 & 0.947 & - \\
 & MS-AQNet & \textbf{0.985} & 0.872 & 0.310 & 0.471 & 0.951 & - \\
 & \textbf{CAA Ensemble} & 0.803 & \textbf{0.997} & \textbf{0.930} & \textbf{0.862} & \textbf{0.995} & - \\ \midrule
 
\multirow{3}{*}{\textbf{Other}} 
 & Baseline CNN & 0.809 & 0.884 & 0.619 & 0.702 & 0.933 & - \\
 & MS-AQNet & 0.908 & 0.820 & 0.596 & 0.719 & 0.897 & - \\
 & \textbf{CAA Ensemble} & \textbf{0.937} & \textbf{0.989} & \textbf{0.962} & \textbf{0.950} & \textbf{0.997} & - \\ \midrule
 
\multicolumn{8}{c}{\textit{Overall Performance (Macro Average)}} \\ \midrule
\multirow{3}{*}{\textbf{All Classes}} 
 & Baseline CNN & 0.677 & 0.898 & 0.594 & 0.612 & 0.896 & 0.553 \\
 & MS-AQNet & 0.670 & 0.876 & 0.522 & 0.533 & 0.866 & 0.443 \\
 & \textbf{CAA Ensemble} & \textbf{0.916} & \textbf{0.979} & \textbf{0.931} & \textbf{0.922} & \textbf{0.996} & \textbf{0.942} \\ 
\midrule \midrule

\multicolumn{8}{c}{\textit{\textbf{B. Plus Disease (Binary Vascular)}}} \\ 
\midrule
\multirow{3}{*}{\textbf{Plus Disease}} 
 & Baseline CNN & 0.778 & 0.962 & 0.594 & 0.674 & 0.977 & 0.647 \\
 & VascuMIL & \textbf{0.975} & 0.971 & 0.712 & 0.823 & 0.995 & 0.808 \\
 & \textbf{CAA Ensemble} & 0.901 & \textbf{0.996} & \textbf{0.936} & \textbf{0.918} & \textbf{0.999} & \textbf{0.897} \\ 
\bottomrule
\end{tabular*}
\begin{tablenotes}
\small
\item \textit{Note: Baseline CNN utilizes EfficientNet-B0 with late fusion by concatenation.}
\end{tablenotes}
\end{threeparttable}
\end{table*}

\subsection{Implementation Details}
The experiments were carried out utilizing PyTorch on a single NVIDIA GPU with mixed precision (AMP).  

\subsubsection{Software Configuration}
The preprocessing pipeline was implemented using \textit{OpenCV} for morphological operations and \textit{Scikit-Image} for vesselness filtering. The texture patches were extracted at a resolution of $224 \times 224$ pixels from the processed resolution of $768 \times 768$ pixel inputs to ensure sufficient receptive field.

\subsubsection{Training Strategy}
We employed a regimen to combat class imbalance. All stages utilized the AdamW optimizer with cosine annealing.

\begin{itemize}
    \item \textbf{Structure Stream (MS-AQNet):} Trained for a total of 30 epochs with a batch size of 16. Warm-up approach has been used: freezing the backbone network for the initial 5 epochs and then full-network training with different learning rates ($10^{-3}$ for heads, $10^{-6}$ for backbone).
    
    \item \textbf{Texture Stream (VascuMIL):} Represented as bags of $N = 24$ patches. Trained for 30 epochs with a bag batch size of 4. Due to Plus Disease data sparsity ($\approx 10.5\%$), a bag-level weighted random sampler with an equal probability of selecting positive and negative bags ($p=0.5$) is used.
    
    \item \textbf{Fusion Ensemble:} Trained via stacking using frozen feature logits from validation folds. The Fusion MLP was trained for 80 epochs with a learning rate of $10^{-2}$ to permit the shallow MLP to learn optimal decision boundaries without overfitting.
\end{itemize}

\subsubsection{Evaluation Metrics}
To assess diagnostic performance, we report standard classification metrics including Sensitivity (Recall), Specificity, Precision, F1-Score, Cohen's Kappa ($\kappa$), and the Area Under the ROC Curve (AUC).

Crucially, model selection (checkpointing) was governed by a custom composite \textit{Clinical Score ($CS$)}. This metric averages the F1-Score with the sensitivity of the target pathology (Severe ROP or Plus Disease), ensuring that the selected model prioritizes safety (high recall) without sacrificing overall stability:

\begin{equation}
CS = \frac{1}{2} \left( F1_{macro} + \text{Sensitivity}_{target} \right)
\label{eq:cs}
\end{equation}

\section{Results}
\label{sec:results}

% --- FIGURE 4 ---
\begin{figure*}[!t]
\centering
\includegraphics[width=\textwidth]{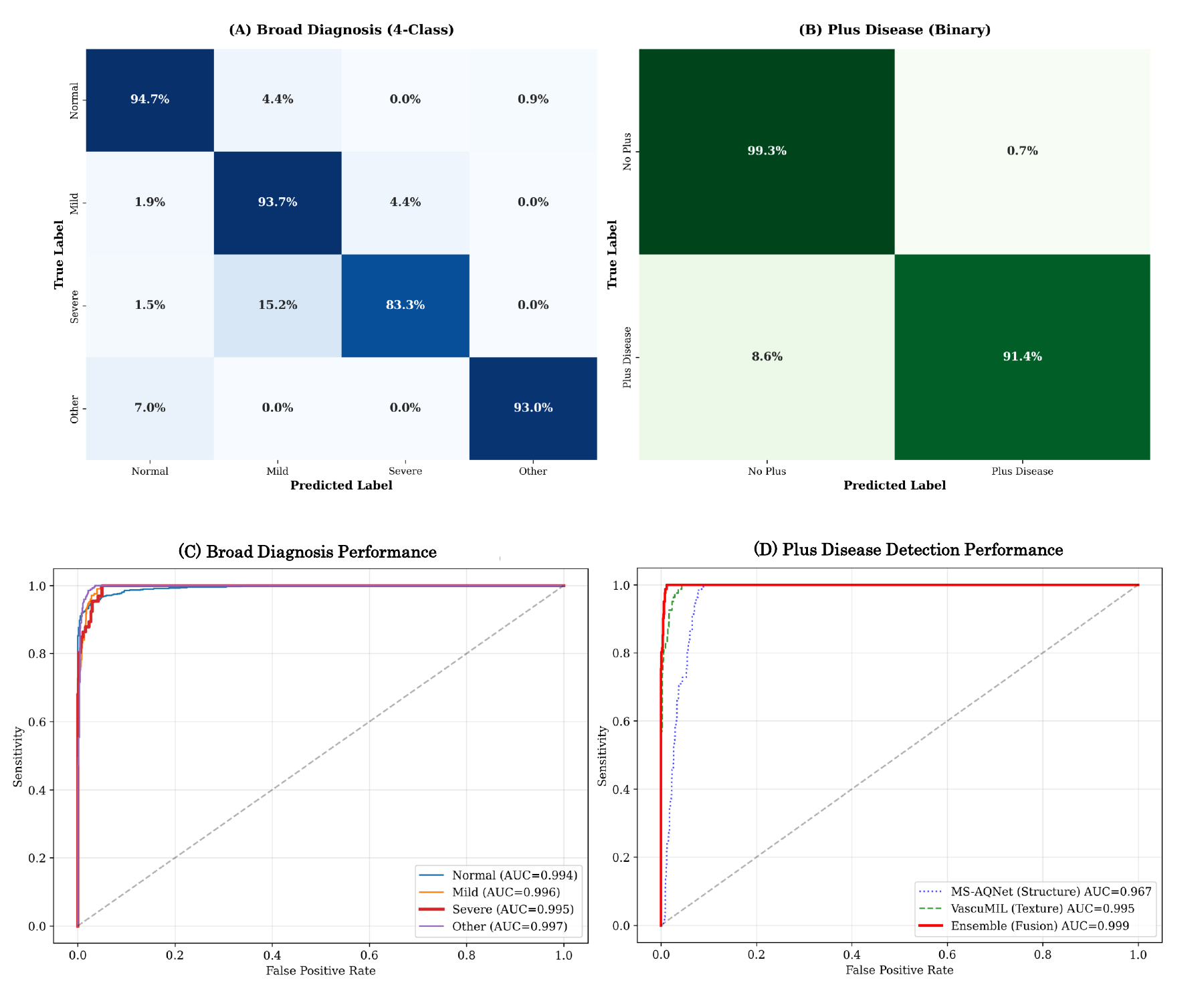} 
\caption{\textbf{Diagnostic Performance Analysis.} (A-B) Normalized confusion matrices for Broad Diagnosis and Plus Disease, demonstrating high diagonal dominance and minimal critical errors. (C) Per-class ROC curves for the Ensemble model, showing robust separation across all diagnostic categories ($AUC > 0.99$). (D) ROC curves for Plus Disease detection, highlighting the superior AUC of the Ensemble (Red) compared to individual structure (Blue) and texture (Green) specialists.}
\label{fig:performance_curves}
\end{figure*}
% ----------------

\subsection{Diagnostic Performance Evaluation}
\label{subsec:diagnostic_assessment}
The primary evaluation is focused on the system's sensitivity and specificity in correctly identifying sight-threatening ROP from benign conditions based on both structural and vascular features. The class-wise performance of the proposed Ensemble compared to the standard Baseline CNN and individual specialist models on the test set is shown in Table \ref{tab:clinical_performance}.

For \textit{Broad Diagnosis} (Task A), the standard Baseline CNN struggled with inter-class ambiguity, yielding only moderate reliability ($\kappa=0.55$). Besides, the baseline employs ``passive fusion,'' extracting features blindly without clinical guidance. In contrast, the proposed MS-AQNet utilized the Active Query mechanism to prioritize physiological validity by ensuring high sensitivity ($0.985$) for severe ROP by spatially gating relevant anatomical regions with a trade-off in precision ($0.310$). The Ensemble model was successful in resolving these trade-offs and achieved a \textit{Macro F1-Score of $0.922$} and a \textit{Cohen's Kappa ($\kappa$) value of $0.942$}, which signifies a level of agreement as almost perfect among the expert annotations and predictions. Furthermore, the model rectified the false positives in the Mild class by enhancing Sensitivity from $0.21$ to $0.97$, which validates the hypothesis that the fusion of the vascular texture modality can reduce ambiguities in the structure.

For \textit{Plus Disease} (Task B), while the Baseline CNN offered reasonable detection capability ($F1=0.67$), the texture expert VascuMIL had superior baseline performance metrics ($AUC=0.995$) but relatively low precision ($0.712$). The proposed CAA Ensemble improved results further, attaining an AUC of $0.999$ and precision of $0.936$ while maintaining high sensitivity. As visible from Fig. \ref{fig:performance_curves}, the confusion matrices show misclassifications are minimal and confined to clinically adjacent categories, with zero critical errors between Normal and Severe classes.

% --- FIGURE 5: SALIENCY ---
\begin{figure*}[!t]
\centering
\includegraphics[width=\textwidth]{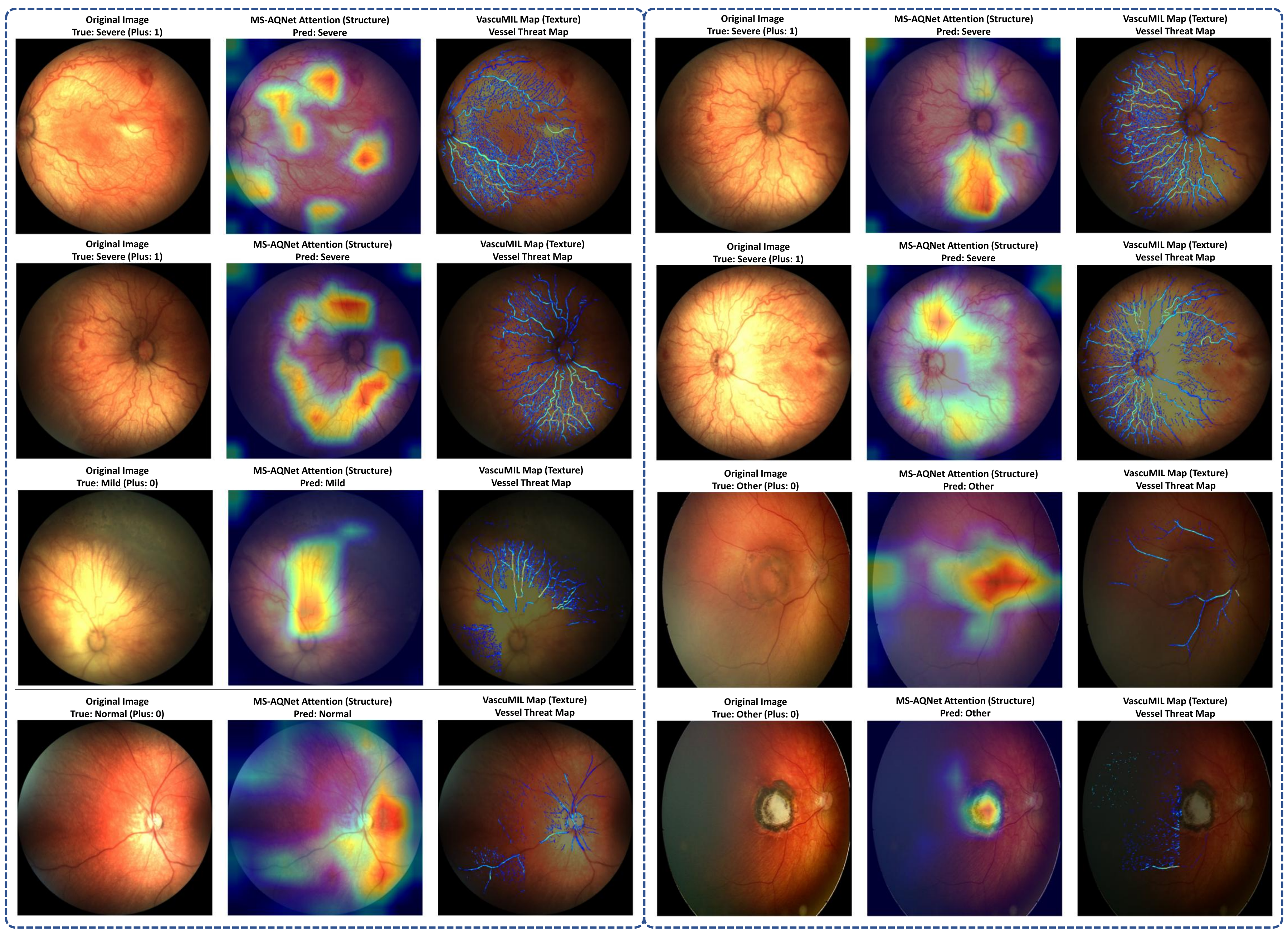}
\caption{\textbf{Multi-Modal Clinical Explainability.} Comparison of diagnostic attention maps. \textbf{(Middle)} MS-AQNet localizes global structural anomalies (e.g., ridges in Row 2). \textbf{(Right)} VascuMIL isolates micro-vascular tortuosity (Red/Yellow) from background (Blue).}
\label{fig:saliency_comparison}
\end{figure*}
% --------------------------

% --- TABLE S.I: BACKBONE (Fixed Overlap using Tabularx) ---
\begin{table}[t]
\centering
\begin{threeparttable}
\caption{Impact of Backbone Architecture on Broad Diagnosis}
\label{tab:backbone_benchmark}
\renewcommand{\arraystretch}{1.3}
% We use a fixed-width 'p' column for the Backbone and Params 
% to stop the overlap with the flexible 'C' columns.
\begin{tabularx}{\columnwidth}{@{}l p{35pt} C C C C C@{}}
\toprule
\textbf{Backbone} & \textbf{Params} & \multirow{2}{*}{\textbf{Sens.}} & \multirow{2}{*}{\textbf{Spec.}} & \multirow{2}{*}{\textbf{Prec.}} & \multirow{2}{*}{\textbf{F1}} & \multirow{2}{*}{\textbf{AUC}} \\ 
\textbf{Architecture} & \textbf{(M)} & & & & & \\
\midrule
\textbf{EfficientNet-B0} & \textbf{7.44} & \textbf{0.91} & \textbf{0.98} & \textbf{0.93} & \textbf{0.92} & \textbf{0.99} \\ 
\midrule
MobileNetV3-L & 8.52 & 0.89 & 0.97 & 0.92 & 0.91 & 0.98 \\
DenseNet-121 & 14.02 & 0.88 & 0.96 & 0.90 & 0.89 & 0.98 \\
ConvNeXt-Tiny & 55.74 & 0.86 & 0.96 & 0.88 & 0.87 & 0.97 \\
ResNet-50 & 47.14 & 0.78 & 0.93 & 0.81 & 0.79 & 0.94 \\
Inception-v3 & 43.68 & 0.76 & 0.92 & 0.79 & 0.77 & 0.93 \\ 
\bottomrule
\end{tabularx}
\begin{tablenotes}
\footnotesize
\item \textit{Note: Performance metrics are rounded to 2 decimals for brevity. `Params' includes the total parameter count of the proposed framework utilizing the specific backbone.}
\end{tablenotes}
\end{threeparttable}
\end{table}
% --------------------------------------------------------

\subsection{Impact of Backbone Architecture}
\label{subsec:backbone_impact}
To validate the selection of the EfficientNet-B0 backbone, we compared architectures ranging from lightweight CNNs to heavy Transformers on the representative Fold 0. As detailed in Table \ref{tab:backbone_benchmark}, a distinct inverse correlation was observed between model capacity and generalization performance on this small dataset ($N=188$).

The large models such as \textit{ResNet-50} ($\sim$47M parameters) and \textit{Inception-v3} ($\sim$43M) were prone to overfitting and performed worst in terms of Macro F1 score (0.74--0.79). Slightly more recent models such \textit{ConvNeXt-Tiny} were more robust (F1=0.875) compared to older residual networks but could not outperform compact models’ efficiency. \textit{EfficientNet-B0} performed best across all metrics with the lowest parameter size (7.44M). This confirms that architectural compactness acts as a vital regularizer by preventing the memorization of patient-specific noise for small dataset.

% --- TABLE 4: ABLATION (Full Width, Auto-Spaced) ---
\begin{table*}[!t]
\centering
\begin{threeparttable}
\caption{Ablation Study: Incremental Impact of Architectural Components}
\label{tab:ablation}
\renewcommand{\arraystretch}{1.2}
\setlength{\tabcolsep}{0pt} % Let fill handle the spacing

\begin{tabular*}{\textwidth}{@{\extracolsep{\fill}} l c c c c c c c c c }
\toprule
\multirow{2}{*}{\textbf{Method Configuration}} & \multicolumn{3}{c}{\textbf{Components}} & \multicolumn{6}{c}{\textbf{Performance Metrics (Mean)}} \\ 
\cmidrule(lr){2-4} \cmidrule(lr){5-10}
 & \textbf{Meta} & \textbf{VMAP} & \textbf{Attn.} & \textbf{Sens.} & \textbf{Spec.} & \textbf{Prec.} & \textbf{F1} & \textbf{Kappa} & \textbf{AUC} \\ 
\midrule

\multicolumn{10}{l}{\textit{Structural Stream (Evaluated on Broad Diagnosis)}} \\
1. Baseline CNN & - & - & - & 0.645 & 0.890 & 0.680 & 0.662 & 0.385 & 0.892 \\ 
2. MS-AQNet (Image) & - & - & Active & 0.720 & 0.915 & 0.745 & 0.730 & 0.412 & 0.935 \\ 
3. MS-AQNet (Full) & \checkmark & - & Active & 0.670 & 0.876 & 0.522 & 0.533 & 0.443 & 0.866 \\ 
\midrule

\multicolumn{10}{l}{\textit{Texture Stream (Evaluated on Plus Disease)}} \\
4. VascuMIL (RGB) & - & - & Gated & 0.910 & 0.935 & 0.620 & 0.735 & 0.710 & 0.965 \\ 
5. VascuMIL (+VMAP) & - & \checkmark & Gated & 0.975 & 0.971 & 0.712 & 0.823 & 0.808 & 0.995 \\ 
\midrule

\multicolumn{10}{l}{\textit{Final System (Evaluated on Broad Diagnosis)}} \\
\textbf{6. Proposed CAA Ensemble} & \textbf{\checkmark} & \textbf{\checkmark} & \textbf{Hybrid} & \textbf{0.919} & \textbf{0.973} & \textbf{0.927} & \textbf{0.926} & \textbf{0.945} & \textbf{0.997} \\ 
\bottomrule
\end{tabular*}
\begin{tablenotes}
\footnotesize
\item \textit{Note: `Meta' = Clinical Metadata injection. `VMAP' = Vascular Topology Maps. `Attn.' = Attention Mechanism. Rows 1-3 and 6 report Macro-Averages for 4-Class Diagnosis. Rows 4-5 report Binary metrics for Plus Disease. To reduce the computational burden, the ablation study were carried out on the \textit{Fold 0}}
\end{tablenotes}
\end{threeparttable}
\end{table*}
% --------------------------------------------------------

% --- FIGURE 6: FEATURE SPACE ---
\begin{figure*}[!t]
\centering
\includegraphics[width=\textwidth]{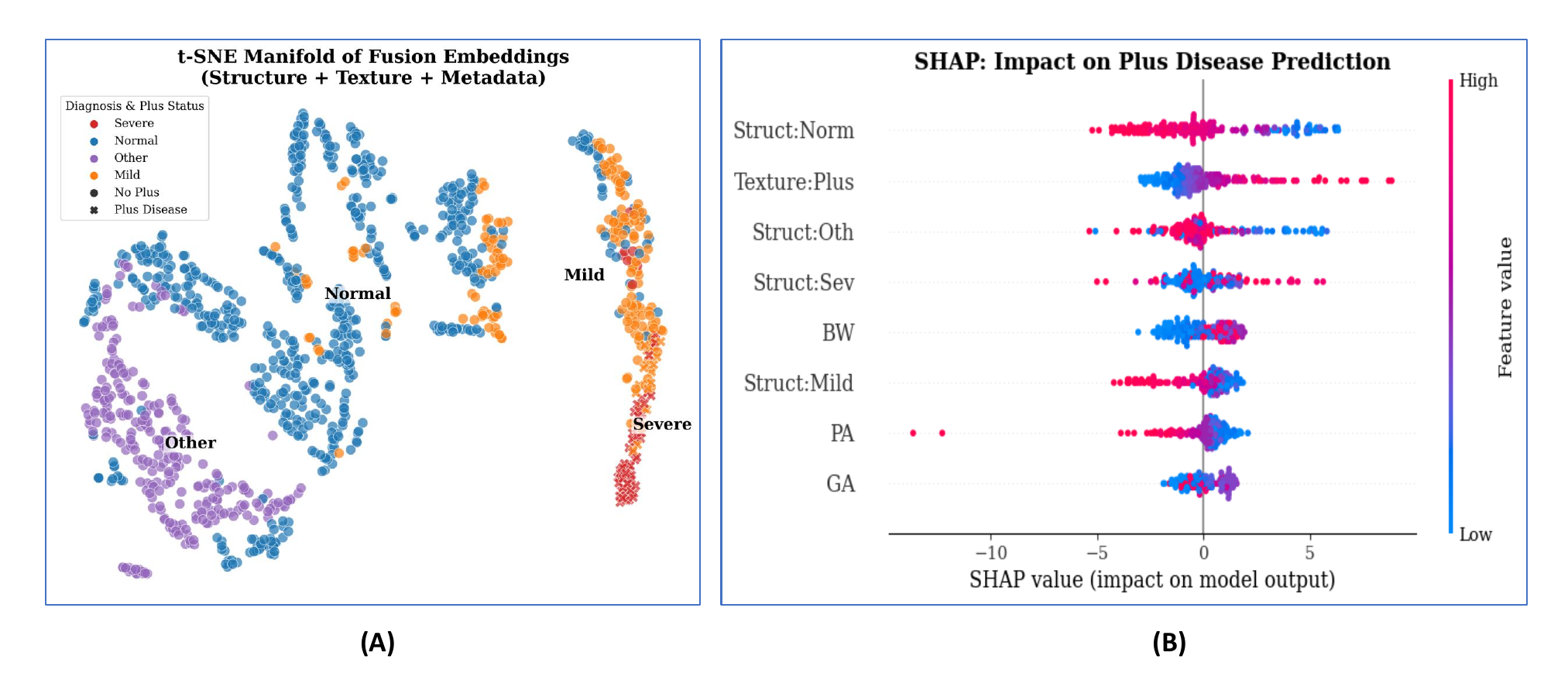}
%trim={10mm 0 0 0},clip, 
\caption{\textbf{Feature Space and Decision Logic.} \textbf{(A) t-SNE Manifold:} Fusion embeddings show a clear progression from Normal to Severe, with Plus Disease cases clustered in high-severity regions. \textbf{(B) SHAP Analysis:} Vascular texture (`Plus') drives prediction, dynamically modulated by structural severity and Gestational Age.}
\label{fig:feature_space}
\end{figure*}
% -------------------------------

\subsection{Ablation Study}
\label{subsec:ablation_study}
We conducted a systematic ablation study (Table \ref{tab:ablation}) to quantify the additive value of the proposed mechanisms.

\textit{Structural Stream (Rows 1--3):}
The baseline CNN model (Row 1) had a low value of sensitivity (0.645), suggesting that the visual features alone cannot be used for accurate staging. Adding the \textit{Active Query} component (Row 2) improved the value of sensitivity to $0.720$. Further addition of metadata injection (Row 3) resulted in high-recall safety that dragged the value of precision down to $0.522$. This validates that clinical contexts modulate visual attention to flag potential pathology.

\textit{Texture Stream (Rows 4--5):}
In the vascular domain, the inclusion of \textit{Vascular Topology Maps (VMAP)} proved to be critical. A four-channel image input (Row 5) stimulated an increased AUC value for the Plus Disease from $0.965$ to $0.995$ and Kappa from $0.71$ to $0.80$ when compared to the RGB-only MIL (Row 4) highlighting the fact that explicit geometric priors are necessary to resolve subtle micro-vascular tortuosity.

\textit{Synergistic Fusion (Row 6):}
The Final Ensemble reached its highest level of reliability ($\kappa=0.945$) and F1-Score ($0.926$). By integrating a high-sensitivity structural stream with a high specificity texture stream, false positives were effectively eliminated by Ensemble model which supports the hypothesis that structural and vascular features provide orthogonal diagnostic signals.

\subsection{Multi-Modal Explainability and Feature Space Analysis}
\label{subsec:explainability}

To validate the `Glass Box' nature of the framework, we analyzed the decision-making process across both the visual domain and the latent feature space.

\subsubsection{Orthogonal Attention Mechanisms}
Fig. \ref{fig:saliency_comparison} compares  the diagnostic foci of the specialist streams. The \textit{MS-AQNet} creates diffuse heatmaps, emphasizing the edge anomalies such as ridges, commonly ignoring the central vasculature in non-plus cases. The \textit{VascuMIL} model creates sharp vascular threat maps which isolates tortuosity signals (visualized in neon) while shadowing the retinal background. This clear separation confirms that the model mimics expert screening by integrating distinct structural and vascular clues.

\subsubsection{Manifold Organization and Decision Logic}
The t-SNE projection of the fusion embeddings (Fig. \ref{fig:feature_space}A) show a reasonable disease course: Mild ROP (Orange) positions between the Normal (Blue) and Severe (Red) clusters. The clusters of the Plus Disease cases (`x') are very close to the high severity regions, confirming the model effectively correlates vascular tortuosity with structural features.

In complement, SHAP analysis (Fig. \ref{fig:feature_space}B) validates the synergistic fusion logic. The texture-based logit emerges as the strongest predictor of vascular disease, with logits and priors from structure (GA) playing a crucial role in moderating it. This combined effect ensures robust decision-making in outlier cases.

\section{Discussion}
\label{sec:discussion}

\subsection{Clinical Utility and Workflow Integration}

The main advantage of an automatic screening system lies in its ability to serve as a safe and scalable triage tool. Our ensemble demonstrates a high alignment with this mandate. In particular, the proposed approach reaches a sensitivity of $0.916$ for Broad Diagnosis and $0.901$ for Plus Disease, hence keeping with safety-first principles as specified in ETROP guidelines and preventing a possible risk of false negatives~\cite{good_final_2004}.

Aside from these quantitative factors, the system supports the concept of task shifting in telemedicine environments. Currently, the problem with screening involves the limited availability of ophthalmologists. The open or glass box framework allows the technician to obtain images and utilize the \textit{Vascular Threat Maps} and \textit{Structural Heatmaps} to detect high-risk infants for evaluation. The interpretability components serve as the quality-control step whereby experts can assess if the result indicates the severe ridge or if there are artifacts instead.

\subsection{The Power of Asymmetric Ensembling and Inductive Bias}
Methodologically, our results challenge the big data paradigm. Benchmarks such as DeepROP~\cite{wang_automated_2018} and i-ROP~\cite{redd_evaluation_2019} depend on brute-force optimization over cohorts exceeding $20,000$ images. In contrast, we observed that heavy architectures like ResNet-50 exhibit feature collapse on small cohorts ($N=188$), as evidenced by a stagnating F1 score of $0.796$.

We demonstrate that \textit{architectural inductive bias} can effectively bridge this data gap. By employing a resolution bifurcation strategy, the optimization problem becomes tractable. Crucially, the integration of clinical context is hierarchical: the \textit{Active Query} mechanism acts as a spatial guide, injecting metadata to constrain the visual search to relevant anatomical zones, while the subsequent \textit{FiLM layer} provides semantic calibration, scaling feature responses based on developmental risk profiles. Simultaneously, the \textit{VascuMIL} stream leverages explicit geometric contexts (VMAP) to isolate vessel topology. This asymmetric synergy increased Cohen’s Kappa from $0.44$ to $0.94$, capturing orthogonal diagnostic signals that single-stream CNNs fundamentally miss.

\subsection{Comparison with State-of-the-Art}
Although direct comparisons are constrained by data heterogeneity, our method compares favorably in data efficiency. DeepROP~\cite{wang_automated_2018} requires over 20 thousand images for $>96\%$ sensitivity, while we achieved a similar level of safety from a dataset two orders of magnitude smaller in our ensemble framework. Furthermore, unlike passive fusion approaches \cite{kang_multimodal_2021}, our Active Query leverages metadata in the feature space. As evidenced by the performance gap against the representative passive baseline (Table \ref{tab:ablation}), this mechanism demonstrates superior generalization abilities. This proves that in the medical `long tail,' intelligent architecture can replace the need for brute-force data scaling.

\subsection{Limitations and Future Directions}
Despite the robust internal validation using `Stratified Group K-Fold,’ the study has the drawback of being a single-center study. External validation within diverse geographical populations is necessary for it to become generalizable, particularly given the fact that different risk factors exist for high-income and middle-income countries ~\cite{hong_retinopathy_2022}. Further, even if it has a very high precision for detecting referable ROP cases, this method has not quantified the stage width or clock hours, which is secondary measurements conventionally made for surgical consideration. Future research should be conducted to analyze changes in vascular topologies over time in order to forecast disease progression before it could lead to sight-threatening diseases.

\section{Conclusion}
\label{sec:conclusion}
This work presents a paradigm shift from black box classification to \textit{Context-Aware Asymmetric Ensembling} for ROP screening. The design combines a clinically gated structural specialist and an anatomy-aware vascular specialist, achieving a state-of-the-art result on the publicly available dataset.

Notably, the model generalizes from binary prediction to offer visual clinical reasoning by the distinction between the severity of structure via attention heatmaps and the risk of blood vessels via topology maps. Our findings demonstrate that we can democratize expert screening simply by forcing the model to follow clinical logic, such as separating texture from structure. This provides a realistic path toward solving the ROP crisis in underserved areas with a system that is both scalable and easy to interpret.

\section*{Code Availability}
To facilitate reproducibility and future research, the complete source code, pretrained model weights, and preprocessing pipelines are publicly available at: \url{https://github.com/mubid-01/MS-AQNet-VascuMIL-for-ROP_pre}.

\section*{CRediT authorship contribution statement}

\textbf{Md. Mehedi Hassan:} Writing – review and editing, Writing – original draft, Visualization, Validation, Software, Methodology, Investigation, Formal analysis, Data curation, Resources, Conceptualization. \textbf{Taufiq Hasan:} Writing – review and editing, Validation, Supervision, Project administration, Investigation, Funding acquisition, Formal analysis.

\section*{Declaration of competing interest}

The authors declare that they have no known competing financial interests or personal relationships that could have appeared to influence the work reported in this paper.

\FloatBarrier
\bibliographystyle{elsarticle-num}
\bibliography{References}
\end{document}